# Quasiparticle Poisoning and Quantum Coherence in a Differential Charge Qubit

J.F. Schneiderman,[1] M.D. Shaw,[1] B. Palmer,[2] P. Delsing,[3,4] and P.M. Echternach[4]

[1] University of Southern California, Dept of Physics and Astronomy, Los Angeles, CA, 90089-0484
[2] Laboratory for Physical Sciences, College Park, MD, 20740
[3] Chalmers University of Technology, Microtechnology and Nanoscience, MC2, 412 96 Göteborg, Sweden
[4] Jet Propulsion Laboratory, California Institute of Technology, Pasadena, CA, 91109

We demonstrate the operation of a differential single Cooper-pair box, a charge qubit consisting of two aluminum islands, isolated from ground, coupled by a pair of small-area Josephson junctions, and read out with a superconducting differential radio-frequency single electron transistor. We have tested four devices, all of which show evidence of quasiparticle poisoning. The devices are characterized with microwave spectroscopy and temperature dependence studies, and Coulomb staircases are shown to be e-periodic in all samples. However, coherent control is still possible with non-adiabatic voltage pulses. Coherent oscillation experiments and a relaxation time measurement were performed using a charge derivative readout technique.

## I. INTRODUCTION

The single Cooper-pair box (SCB) is a mesoscopic electronic circuit which behaves in certain limits as a two-level quantum system, or qubit.[1,2,3] In recent years, there has been a sustained effort to develop the SCB charge qubit into a practical and reliable building block for quantum computation.[4,5,6] The SCB consists of a small metal island coupled to a large grounded lead via a low-capacitance Josephson junction. By applying a voltage to a small gate capacitor, individual Cooper-pairs can tunnel coherently, and arbitrary superpositions of charge states can be reliably prepared.[7,8,9] For temperatures and Josephson coupling energies much lower than the charging energy, the energy bands of the SCB effectively reduce to two levels. The system can be described by the Hamiltonian $H = -\frac{1}{2}\left(4E_c(1-n_g)\sigma_z + E_J\sigma_x\right)$ in the basis of Cooper-pair number states, where $E_c = \frac{e^2}{2(C_G+C_J)}$ is the single-electron charging energy, $C_{G,J}$ are the gate and junction capacitances, respectively, $n_g = \frac{C_g V_g}{e}$ is the reduced gate charge in electrons, $\sigma_{x,z}$ are the Pauli matrices, and the Josephson coupling energy $E_J$ is proportional to the critical current of the junction.

In the present work, we have experimentally investigated an alternative charge qubit layout, the differential single Cooper-pair box

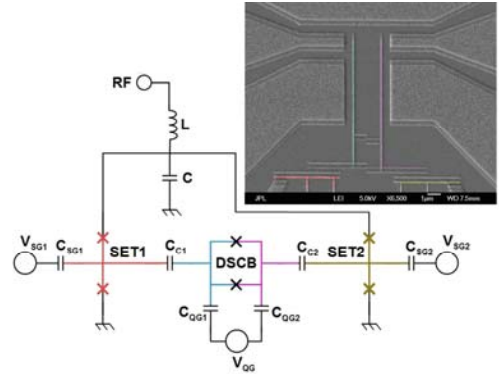

Fig. 1. (color online) Circuit diagram showing the DSCB and DRFSET readout. Note that the two SSETs are read out in parallel with a single tank circuit. $C_{QG1}$, $C_{QG2} \sim 30$ aF are the qubit gate capacitances, $C_{SG1}$, $C_{SG2} \sim 110$ aF are the SET gate capacitances, and $C_{C1}$, $C_{C2} \sim 75$ aF are the island-SSET coupling capacitances. The junction capacitances are $\sim 820$ aF. *Inset:* SEM image of sample 1. The red and yellow features are the two SSETs, while the violet and cyan features are the DSCB islands. The DC-SQUID loop is in the lower center.

(DSCB).[10,11,12,13] In contrast to the conventional SCB, the DSCB consists of two small superconducting islands coupled by low-capacitance Josephson junctions. No reservoirs are present, leaving the entire structure isolated from ground. The potential of each island is controlled independently with separate gate capacitors, as shown in Figure 1. As the differential gate charge $n_g = \frac{1}{2}(n_{g2} - n_{g1})$ is increased across the DSCB gates, single Cooper-pairs can coherently tunnel from one island to the other. With the substitution of charge difference states $n = \frac{1}{2}(n_2 - n_1)$ for charge number states, the Hamiltonian for the DSCB is identical to that of the conventional SCB.[10] In the two-level approximation, the computational basis of the differential qubit is $|0\rangle \to |10\rangle$, $|1\rangle \to |01\rangle$ where $|n_1 n_2\rangle$ is the charge number state with $n_1$ ($n_2$) excess Cooper-pairs on the left (right) qubit island. The differential charge states are read out with a superconducting differential radio-frequency single-electron transistor (DRFSET), which consists of a pair of superconducting single electron transistors (SSETs) read out in differential mode with a single LC tank circuit.[11,14]

## II. PREDICTIONS

The DSCB was predicted to have two principal advantages to the conventional SCB. First, since in the two-level approximation the charge number states on each island are perfectly anticorrelated, the DSCB behaves as a charge qubit embedded in a decoherence-free subspace, which is immune to dephasing from perfectly correlated noise.[11,15] One of the principal sources of decoherence in charge qubits is fluctuating qubit gate voltage. A significant source of these voltage fluctuations is noise from the input of the cold amplifier. As the two SSETs are read out with a single tank circuit and RF amplifier (see fig. 1), the noise from the amplifier at each gate will be strongly correlated in time, limiting its dephasing effect on differential charge states. Another important decoherence mechanism is 1/f noise, one source of which is two-level charge fluctuators embedded in the device substrate, leads, and tunnel barriers.[16,17] For those fluctuators which are located far from the DSCB islands, the potential of both islands will fluctuate together, leaving the differential charge states unaffected. Of course, not all noise sources affecting the DSCB are highly correlated. A significant source is electromagnetic noise coupled capacitively from the SET islands to the qubit islands. Since the two SETs are independent, this noise will be largely uncorrelated. Furthermore, the noise from two-level fluctuators located close to the islands will not have strong spatial correlation on the length scale of the islands, and it is these nearby fluctuators which dominate.

Secondly, the DSCB was predicted to provide a solution to the problem of quasiparticle (QP) poisoning, where QPs tunnel incoherently across the barrier even at very low temperature, where the probability of thermal excitation is exponentially suppressed. QP poisoning poses difficulties for control of the qubit, and has been shown to have adverse effects on qubit coherence times.[18] In particular, it prevents operation of the qubit at the Cooper-pair degeneracy point, where to first order the qubit is immune to dephasing from low-frequency noise, and decoherence times are typically limited by relaxation processes.[5] In the conventional SCB, this problem can be understood by considering non-equilibrium QPs generated in the leads tunneling onto the qubit island.[19,20,21] In the conventional SCB, the lead volume is much larger than the island volume, so the probability that nonequilibrium QPs are generated in the island itself is negligible by comparison. Since the double-island structure of the DSCB has no reservoir and is completely isolated from ground, at zero temperature the parity of the box should depend only on whether the total number of conduction electrons on the box is even or odd,[20] and one would expect that on average, half of the samples fabricated would show evidence of QP poisoning.

## III. EXPERIMENTAL RESULTS

Surprisingly, we fabricated and tested four DSCB samples, and all were strictly poisoned. While the simplest method of determining the parity of each device is to drive the island material normal with a magnetic field, technical limitations (a lack of space inside the helium dewar) made this impractical. Instead, we base this conclusion on spectroscopic data and temperature dependence studies. While it is possible that all four devices started with an odd number of electrons, in hindsight it appears more likely that non-equilibrium QPs generated in the islands themselves could be responsible for QP poisoning at low temperatures. In the conventional grounded SCB, techniques exist to reduce the effect of poisoning by engineering the energy gap difference $\delta\Delta = \Delta_I - \Delta_R$ between the island and the reservoir.[19,22] In such schemes, the energy barrier $\delta\Delta$ prevents QPs generated in the reservoir from tunneling onto the qubit island. In the differential SCB, there is no reservoir, and the quantum number of interest is the charge difference between the two islands. As such, the DSCB requires symmetric construction, where the gap energies of the two islands are almost identical, and such techniques cannot be applied. Note that since the gap energies are almost identical, the odd-state Coulomb staircase should be effectively e-

periodic, i.e. the even-$n$ and odd-$n$ steps will have the same width.[23,24]

### A. Experimental setup

The DSCB devices, an example of which is shown in Fig. 1, were fabricated using a conventional shadow-mask evaporation technique.[25] The devices consist of two isolated aluminum islands (volume ~0.15 μm$^3$ each) coupled by a pair of small-area Al/AlOx/Al tunnel junctions in a DC-SQUID configuration, to allow a tunable Josephson coupling energy $E_J$. Each sample was mounted on the mixing chamber of a dilution refrigerator with a base temperature of 10 mK. Readout was performed using a superconducting differential RF-SET, as described in an earlier work.[14] The resonant frequency of the RF tank circuit was ~620 MHz, and the bandwidth was ~8 MHz. Sensitivities of SSET1 and SSET2 for sample 1 were measured as in Ref. 26 to be $3.5 \times 10^{-4} e/\sqrt{Hz}$ and $2.7 \times 10^{-4} e/\sqrt{Hz}$, respectively. For sample 1, the charging energy $E_C^{SET}/k_b$ = 1.7 K and the superconducting gap energy $\Delta/k_b$ = 2.4 K were extracted from maps of reflected power versus gate voltage and drain-source voltage for both SSET1 and SSET2. Samples 1-3 were of very similar design, while sample 4 was fabricated with small QP traps ($T_c \sim 0.5$ K) in contact with the qubit islands themselves. Relevant device parameters for samples 1-3 are listed in Table 1, with $E_c$ and $E_J$ extracted from microwave spectroscopy. The gate capacitance $C_G$ is computed by comparing the applied gate voltage to the resulting Coulomb staircase, assuming that the staircase is e-periodic. The junction capacitance is then $C_J = \frac{e^2}{2E_C} - C_G - C_C$, where $C_C$ = 75 aF is the coupling capacitance between the qubit and SSET islands. For sample 4, noise was so severe that accurate spectroscopy and temperature dependence data could not be obtained.

Differential Coulomb staircases were measured for all four samples by applying a differential voltage ramp to both qubit gates with a slow (261 Hz or 1200 e/s) ramp frequency. To keep the DRFSET in differential mode, compensating voltage ramps were simultaneously applied to both DRFSET gates. Since the various cross- and gate capacitances are difficult to accurately calculate numerically to within a factor of two, the parity of the device is not readily apparent from the staircase alone. However, it is clear from temperature dependence and spectroscopy data that all four devices are strictly poisoned.[27]

### B. Temperature dependence

Figure 2 shows the results of a temperature-dependence study performed on sample 3. Figure 2a shows Coulomb staircases measured at a range of mixing chamber temperatures from 33 to 336 mK. Note that the periodicity of the staircases remains constant throughout the entire range of temperatures, and that no new features emerge. If the device were not poisoned, the staircase would become completely e-periodic at a temperature $T^* = \frac{\Delta/k_b}{\ln(N_{eff})}$ ~170 mK,[28] where the even-odd free energy difference goes to zero. In this formula, $\Delta/k_b$ = 2.4 K is the superconducting energy gap, $N_{eff} = \rho_0 \Omega \sqrt{2\pi \Delta T^*}$ is the effective number of QP states available for excitation, $\rho_0 = 2.0 \times 10^{24}$ K$^{-1}$m$^{-3}$ is the normal-metal density of states at the Fermi level, and $\Omega = 0.31\,\mu m^3$ is the combined volume of both islands. Figure 2b shows the slope of each staircase at the degeneracy point, plotted with theoretical predictions assuming a thermal population of QP states in the two-level approximation,[29] which is further evidence that the devices are poisoned. In these curves, $E_c/k_B = 0.62$ K and $E_J/k_B = 0.6$ K, values that are in good agreement with those extracted from spectroscopy. The deviation from theory at low temperature is likely caused by electron heating above the lattice temperature due to the decrease of electron-phonon scattering at low temperatures.[30,31]

| Sample | $E_C/k_B$ | $E_J/k_B$ | $C_G$ | $C_J$ |
|---|---|---|---|---|
| 1 | 0.60±0.05K | 0.6 ± 0.3 K | 31 aF | 1.44 fF |
| 2 | 0.60±0.05K | 0.7 ± 0.3 K | 32 aF | 1.44 fF |
| 3 | 0.56±0.05K | 0.6 ± 0.3 K | 30 aF | 1.55 fF |

*Table 1.* Relevant sample parameters extracted from microwave spectroscopy (see Fig. 3). $C_G$ and $C_J$ are the lumped gate and junction capacitances. Sample 4, which was fabricated with quasiparticle traps in direct contact with the qubit islands, could not be reliably tested.

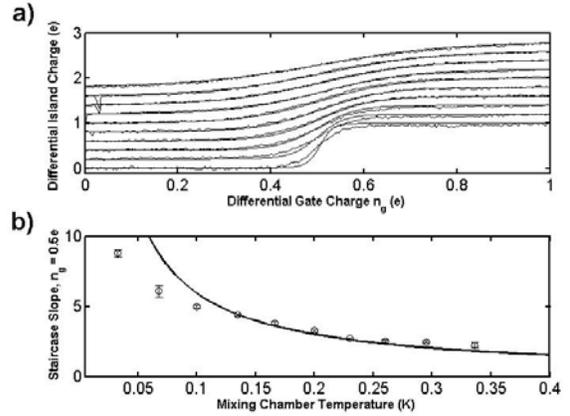

Fig. 2. Temperature dependence of the Coulomb staircase. *a)* Coulomb staircases measured with sample 3 at temperatures ranging (bottom to top) from 33 to 336 mK, shifted vertically for clarity. Note the absence of new features as the temperature is increased. Theoretical fits assume a thermal population of quasiparticle states in the two-level approximation. *b)* Open circles are the slopes of the above staircases at the quasiparticle tunneling point, $n_g = 0.5e$. Deviation from theory at low temperature is likely due to heating from electron-phonon scattering. Sample parameters for the theoretical predictions are $E_C/k_B$ = 0.62 K and $E_J/k_B$ = 0.6 K.

At low temperatures $T \ll T^*$, QP poisoning can be understood in terms of SSET backaction, in the sense that the periodicity of Coulomb staircases in the conventional SCB is dependent on the drain-source voltage of the SSET.[32,33] We have measured staircases at a variety of SSET drain-source voltage biases, including the double Josephson quasiparticle peak (DJQP), the Josephson quasiparticle peak (JQP), and at the edge of the gap, without any apparent change in staircase periodicity. The width of each step remains unchanged, and no new qualitative features emerge. We thus conclude that the DSCB is fully poisoned even when the SSET is biased at the DJQP. The data presented in Figures 2 and 3 were taken with the SET biased at the JQP.

**C. Microwave spectroscopy**

Spectroscopic measurements also showed evidence of QP poisoning. Continuous wave microwaves in the range of 28-40 GHz were added to the voltage ramps with bias tees at low temperature. The microwaves were applied to each qubit gate with a 180° phase difference between them to mimic the effect of cw microwaves in a conventional SCB. Transitions occur when the applied microwave energy is resonant with the DSCB energy level spacing. Spectroscopic data from sample 3 is shown in Fig. 3. Note that resonant transitions occur near the even-odd level crossing at $n_g = 0.5$, and not at the Cooper-pair degeneracy point at $n_g = 1$, implying that the step measured in the Coulomb staircase is due to incoherent QP tunneling, rather than the coherent tunneling of Cooper-pairs. Estimates of the charging energy $E_C$ and the Josephson coupling energy $E_J$ extracted from spectroscopy are listed in Table 1.

**D. Coherent oscillation**

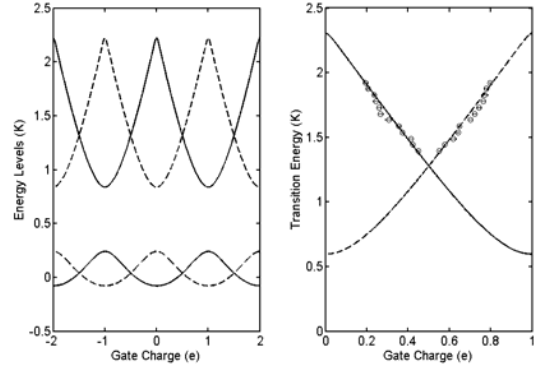

Fig. 3. Spectroscopy of DSCB energy levels. *Left:* DSCB energy bands in the two-level approximation. Solid (dashed) curves are energy eigenvalues for even (odd) parity states. *Right:* Transition energies near the quasiparticle tunneling point. The solid (dashed) line shows the Cooper-pair transition energy for the even (odd) states. Open circles are experimental data with sample 3 showing spectroscopy peak locations as a function of applied microwave energy. This and similar data were used to extract the sample parameters listed in table 1. In this figure, $E_c/k_B = 0.56 \pm 0.05$ K, and $E_J/k_B = 0.6 \pm 0.3$ K. The mixing chamber temperature was 12 mK.

Despite the problem of QP poisoning, coherent control of the DSCB was demonstrated by applying fast DC gate pulses, a familiar technique for state manipulation in charge qubits.[7] In such experiments, a short gate voltage excursion of duration $\Delta t$ is made non-adiabatically from the readout point $n_g = n_{g0}$ to the degeneracy point $n_g = 1$, where the system undergoes coherent oscillation between the two charge states for the duration of the pulse. By varying $\Delta t$, coherent oscillations can be observed. To be non-adiabatic, the rise time of the pulses must be shorter than $h/E_J \sim 70$ ps. Our experiments were performed with an Advantest D3186 pulse generator, with a rise time of 30 ps. While QP tunneling makes the Cooper-pair degeneracy point impossible to reach adiabatically (see Fig. 3), the QP tunneling rates have been observed in the conventional SCB to be on the order of tens of kilohertz.[34,35,36] In our experiments, the excursion time $\Delta t$ is on the order of nanoseconds, so QP tunneling events that occur during the excursion are rare.

Figure 4 illustrates coherent oscillation in the DSCB of sample 1. Note that this is a different sample than that used in the spectroscopy and temperature dependence studies shown in Figs. 2 and 3 (sample 3). In these experiments, the DRFSET is used to measure the derivative of the island charge with respect to gate voltage by applying a small-amplitude (0.04 e) AC excitation ($\omega_{ex}$ = 500 kHz) to the qubit gate and measuring the component of the DRFSET signal at $\omega_{ex}$ with a lock-in amplifier. By reading out the charge derivative instead of the Coulomb staircase, one can in principle operate the qubit directly at the degeneracy point, where dephasing times are longest, although the bandwidth of the DRFSET (~8 MHz) enforces a fundamental limitation on the speed of such a measurement. Such a readout technique is in the spirit of recent direct measurements of the quantum capacitance of the single cooper-pair transistor[37] and single cooper-pair box[38] by RF reflectometry of an LC oscillator. A further advantage of this technique is an improved signal-to-noise ratio (SNR), since the state information is encoded at the frequency $\omega_{ex}$, above the 1/f noise knee. To demonstrate this, we read out a Coulomb charge staircase at $\omega_{ex}$ = 100 kHz, and compared it with a staircase read out at zero frequency. The SNR of the staircase read out at $\omega_{ex}$ was 1.8 times greater than for a staircase read out at DC.

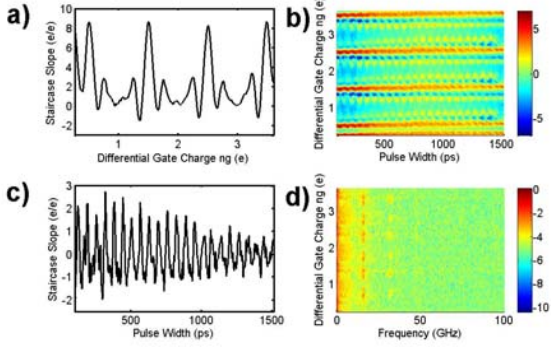

Fig. 4. (color online) Coherent oscillation with DC pulses, measured with sample 1. ***a)***: Representative charge derivative trace (see text). ***b)***: Raw coherent oscillation data with DC pulses. Pulse amplitude is 0.2 e, repetition time 100 ns, pulse rise time 30 ps. Z axis is the staircase slope plotted in units of e/e. ***c)***: Representative slice taken through $n_g$ = 2.36 e for the data shown in (**b**). The mean of each charge derivative trace has been subtracted to compensate for DC drift. ***d)***: Fourier transform of coherent oscillation data shown in (**b**), plotted on a logarithmic scale. Note the peak at 16 GHz, corresponding to $E_J/k_b$ = 0.77 K. The z-axis is a logarithmic plot of the oscillation spectral density in units of $\log_{10} Hz^{-1/2}$.

Figure 4a shows a typical charge derivative signal measured with the DRFSET, which corresponds to the first derivative of the Coulomb staircase. The central peaks at half-integer gate voltages are the derivatives of the steps in an e-periodic Coulomb staircase, while the features to the left and right of these peaks correspond to the charge peaks and dips induced by the DC pulses. The coherent oscillation data shown in Fig. 4b was measured using the charge derivative readout method. Differential DC pulses with width $\Delta t$ varying from 100-1500 ps, a fixed amplitude of ~0.2 e, and a fixed repetition time of 100 ns were applied to the qubit gates through bias tees. Figure 4c shows a representative slice through this data set, with the mean of each charge derivative trace subtracted to eliminate the DC drift in the lock-in amplifier. Figure 4d shows a Fourier transform of the above data, with a clear peak at 16 GHz, corresponding to $E_J/k_B$ = 0.77 K, in good agreement with the value extracted from microwave spectroscopy (see Table 1).

### E. Relaxation time

Despite the presence of QP poisoning, we were also able to measure the relaxation time of the qubit using non-adiabatic gate pulses. As described in earlier works,[39] the qubit is prepared in an excited state with a $\pi/2$-pulse and allowed to decay. As the time between pulses increases, the excited-state contribution to the average charge is reduced. By varying the pulse repetition time $T_R$, the relaxation time $T_1$ can be extracted from the average charge by fitting the decay of the excited state peak to the formula

$$\langle n \rangle = 2n_0 \tanh \frac{T_R}{2T_1} \qquad (1)$$

where $n_0$ is the average charge on the box immediately after applying the pulse.[39]

Figure 5 displays data from a relaxation time measurement performed on sample 1 with the charge derivative readout technique. The data was measured with a fixed pulse width of 300 ps, pulse amplitude ~0.2 e, and a variable repetition time of 90-1000 ns. To compute the average charge on the island at the measurement point, we integrated the peak and dip in the derivative signal for each value of the repetition time. Fits of the data to Eq. (1) yields $T_1$ = 83 ns and $n_0$ = 0.67. This relaxation time is shorter than that measured for a conventional device with similar parameters,[39] although still in the same order of magnitude. While the symmetric layout of the DSCB is predicted to give a longer dephasing time, we would not necessarily expect the relaxation time to be any longer than that of a conventional SCB. While QP tunneling has been shown to have an adverse affect on qubit relaxation times,[18] the measured value of $T_1$ is two orders of magnitude shorter than the estimated QP tunneling times. It is therefore unlikely that $T_1$ is is limited by QP poisoning.

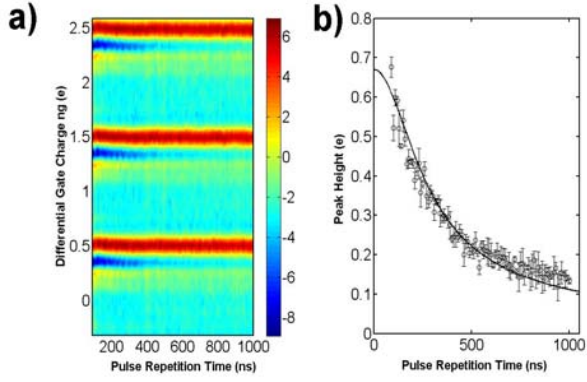

Fig. 5. (color online) Measurement of the qubit relaxation time in sample 1, using non-adiabatic DC pulses. *a)* Raw data, measured using charge derivative method discussed in the text. Pulse width is 300 ps, pulse amplitude is 0.2 e, and repetition time is varied from 90-1000 ns. Z-axis is staircase slope plotted in units of e/e. *b)* Height of the pulse-induced charge peak in the coulomb staircase, obtained by integrating the peaks and dips in the charge derivative signal shown in *a)*. The data is fit to Eq. (1), giving $T_1$ = 83 ns and $n_0$ = 0.67 e.

From Fermi's golden rule, the qubit relaxation rate at zero temperature due to voltage fluctuations coupled capacitively from the qubit gates and SSET islands is given to first order by

$$\Gamma_R = \frac{1}{T_1} = \left(\frac{e(C_c + C_g)E_J}{\hbar C_\Sigma E_0}\right)^2 S_V(E_0/\hbar) \quad (2)$$

where $C_g$ = 30 aF is the qubit gate capacitance, $C_c$ = 75 aF is the coupling capacitance between each SET island and its respective qubit island, $C_\Sigma$ = 1.5 fF is the overall qubit capacitance, $E_0/k_b = \sqrt{(4E_C(1-n_g))^2 + E_J^2}/k_b$ = 1.7 K is the energy level spacing at the gate voltage $n_g$ = 0.3, and $S_V(E_0/\hbar)$ is the spectral density of noise at the frequency commensurate with the level spacing (35 GHz). Using $T_1$ = 83 ns, we find the combined spectral density of voltage noise at the qubit islands coupled capacitively from both SET islands and from both qubit gates to be $1.0 \times 10^{-2}$ nV$^2$/Hz.

For comparison, we consider the expected spectral density of voltage noise coupled from each SET at the DJQP, based on the full quantum mechanical calculation of Clerk et al.[40] Substituting our SSET parameters ($E_C^{SET}/k_b$ = 1.7 K, $E_J^{SET}/k_b$ = 0.16 K, $\Delta/k_b$ = 2.4 K, $R_N$ = 111 k$\Omega$) into eq. 5 of ref. [40] for an SSET biased exactly on the DJQP resonance gives a voltage spectral density of $6.5 \times 10^{-2}$ nV$^2$/Hz on each SSET island and $3.2 \times 10^{-3}$ nV$^2$/Hz at the qubit islands. This theoretical estimate is a factor of three lower than the experimental value extracted from $T_1$, but is in the same order of magnitude, suggesting that noise from the SSET islands is a significant relaxation mechanism in the DSCB. The short relaxation time can presumably be increased by decreasing the SSET-qubit coupling capacitance, using a single-shot DRFSET readout, or using a nondissipative readout technique such as a quantum capacitance measurement.[37,38] In the frequency range commensurate with the qubit energy level spacing and above, Clerk's expression for the spectral density of noise at the SSET island takes the approximate form

$$S_V(\omega) = \left(\frac{eE_J^{SET}}{C_\Sigma^{SET}}\right)^2 \frac{1}{3E_C^{SET}\alpha} \frac{(3E_C^{SET}+\omega)}{(3E_C^{SET}-\omega)\omega^2} \quad (3)$$

where $\alpha = \hbar/e^2 R_N$, $R_N$ is the normal-state tunneling resistance of both SSET junctions in series, $C_\Sigma^{SET} = e^2/2E_C^{SET}$ = 550 aF is the overall capacitance of each DRFSET island, and we have ignored terms of order $\alpha^2$.

### F. Quasiparticle traps

In an attempt to suppress QP poisoning in the DSCB, sample 4 was fabricated with QP traps in direct contact with the qubit islands. The traps were made of an Al-Ti-Au trilayer, which, when deposited underneath the qubit islands, has a superconducting gap energy $\Delta/k_B$ of ~0.9 K.[41] With a smaller superconducting energy gap than the aluminum qubit islands ($\Delta/k_B$ = 2.1 K), the traps were intended to provide a low energy potential well into which QP excitations on the islands could relax. In practice, however, staircases measured from this sample were also e-periodic.

### IV. DISCUSSION

One possible interpretation of these results is that QP poisoning in the DSCB arises due to QP excitation on the islands themselves. In a conventional grounded qubit, the volume of the leads greatly exceeds the volume of the island. As such, previous models[19,20,21] assume that non-equilibrium QPs are generated in the leads and tunnel back and forth between the leads and the island, and that the QP generation rate on the island is negligible by comparison. In the

DSCB, however, the qubit islands are isolated from ground, so QPs generated on the islands will remain there until they recombine. If the QP generation rate on the island is greater than or equal to the recombination rate, then QPs will always be available to tunnel and staircases will be strictly e-periodic at low temperature.

One potential mechanism for generation of non-equilibrium QPs in the qubit islands is Cooper-pair breaking due to noise and radiation from the DRFSET. As current flows through each SSET, fluctuations in the SSET island voltage are capacitively coupled to the qubit island, which have been shown to affect the width and asymmetry of measured Coulomb staircases.[42] Capacitively coupled voltage frequency components which are above the superconducting gap of aluminum can in principle break Cooper-pairs on the qubit islands, as can narrow-band Josephson radiation from the SSET island junctions, or phonons which propagate from the SSET to the qubit island through the substrate.

However, theoretical descriptions of the noise spectrum for the normal-state and superconducting SET do not suggest that capacitively coupled voltage fluctuations alone are capable of explaining nonequilibrium QP poisoning. From Fermi's golden rule, the QP generation and recombination rates on the qubit islands due to voltage fluctuations is given to first order by

$$\Gamma_\pm = \frac{2\pi e^2 \Omega}{\hbar} \int_{\Delta/\hbar}^{\infty} S_V(\pm\omega)\rho(\omega)d\omega \quad (4)$$

where $\Gamma_+$ and $\Gamma_-$ are the rates for Cooper-pair recombination and breaking, respectively, $\rho(\omega) = \frac{k_b \rho_0 |\hbar\omega|}{\sqrt{(\hbar\omega)^2 - \Delta^2}}$ is the density of states per unit volume, $\rho_0$ is the normal-metal density of states at the Fermi level for aluminum, $\Omega$ is the combined volume of the qubit islands, and $S_V(\omega)$ is the power spectral density of voltage noise at the qubit islands themselves. In the sequential tunneling regime of the normal-state SET, the spectral density of voltage fluctuations contains negligible power at large negative frequencies in the low-temperature limit,[43] and hence negligible pair breaking. In the superconducting SET, significant power can be found at negative frequencies, but under most operating conditions the spectral density of noise at high positive frequencies significantly exceeds that at high negative frequencies, so the QP relaxation rate will exceed the excitation rate and the steady-state QP population will be zero. A notable exception is the SSET tuned just off of the DJQP resonance, where the spectral density of noise at negative frequencies is greater than that at positive frequencies, and a qubit population inversion has been predicted.[40] However, since QP poisoning is a significant problem at all SSET bias voltages, it cannot be fully explained by this mechanism alone. Alternatively, narrow-band radiation from an SSET at frequencies extending well above the superconducting gap energy has recently been measured directly,[44,45] and may present a mechanism for QP excitation on the qubit islands. Another possible mechanism may be the propagation of phonons from the SSET to the qubit islands through the substrate.

The hypothesis of nonequibrium QP generation on the qubit island from an external energy source such as the SSET is consistent with our observation of severe QP poisoning and poor performance in the sample fabricated with QP traps in direct contact with the qubit islands. In this device, the superconducting gap energy of the traps is substantially lower than that of the aluminum islands, and in the presence of noise, electromagnetic radiation, or phonons, the traps can become a non-equilibrium QP source. This conclusion is also consistent with recent observations that the severity of QP poisoning is dependent on SSET drain-source bias voltage.[32]

## V. CONCLUSIONS

To summarize, we have fabricated and tested four isolated DSCB circuits, all of which suffer from QP poisoning despite their predicted advantages. We have demonstrated that these devices are poisoned by considering the temperature dependence of the slope of Coulomb staircases and from spectroscopic analysis. No short step appears when the mixing chamber temperature is raised above *T\** or when the SET is biased at the edge of the gap. While poisoning in the DSCB is a major obstacle in the development of such devices as superconducting qubits, coherent control of the DSCB is still possible with non-adiabatic DC pulses, since this manipulation takes place on a shorter time scale than that associated with QP tunneling. We have demonstrated such control by observing coherent oscillations and measuring the qubit relaxation time.

While the differential qubit has several proposed advantages, significant development is required to match the performance of the conventional single Cooper-pair box. A natural improvement is to move away from dissipative readout schemes. Recent experiments performed in our laboratory and elsewhere[38] which measure the quantum capacitance of a conventional SCB with a resonant LC circuit show a complete absence of QP poisoning at low temperatures. However, it is not clear that the DSCB can be isolated well enough from its environment to completely solve the problem of QP poisoning. Furthermore, the resources required to operate a DSCB are almost equivalent to those required to operate two coupled conventional qubits, which can always be operated as a single qubit in differential mode.[15]

## ACKNOWLEDGEMENTS

We would like to thank Rich E. Muller for performing the electron-beam lithography. We would also like to thank Hans Bozler, Keith Schwab, Göran Johansson, and Alexander Korotkov for helpful discussions and advice. This work was performed at the Jet Propulsion Laboratory, California Institute of Technology, under a contract with the National Aeronautics and Space Administration (NASA) and was supported in part by the National Security Agency (NSA).